\documentstyle[aps,epsfig]{revtex}              

\begin{document}
\title{Real null coframes in general relativity and GPS type  coordinates}

\author{M.\ Blagojevi\'c\footnote{On leave from: Institute of Physics,
      P.O.\ Box 57, 11001 Belgrade, Yugoslavia; email: mb@phy.bg.ac.yu}, 
      J.\ Garecki\footnote{On leave from: Institute of Physics, University 
      of Szczecin, 70-451 Szczecin, Poland; 
      email: garecki@wmf.univ.szczecin.pl}, 
      F.W.\ Hehl\footnote{email: hehl@thp.uni-koeln.de}, 
      Yu.N.\ Obukhov\footnote{On leave from: Department of Theoretical 
      Physics, Moscow State University, 117234 Moscow, Russia; \\
      email: yo@thp.uni-koeln.de }} 
\address{Institute for Theoretical  Physics, 
         University of Cologne, 50923 K\"oln, Germany}

\date{27 November 2001}
\maketitle

\begin{abstract}
  Based on work of Derrick, Coll, and Morales, we define a `symmetric'
  null coframe with {\it four real null covectors}. We show that this
  coframe is closely related to the GPS type coordinates recently
  introduced by Rovelli\cite{Rovelli}.
\end{abstract}

 
\section{Introduction}

In the pseudo-Riemannian spacetime of general relativity theory, we
can introduce at each point of spacetime a local vector basis or frame 
$e_\alpha(x)$, with $\alpha,\beta,\dots=0,1,2,3$. As already indicated
by the notation, the zeroth leg $e_0$ is conventionally chosen to be
timelike whereas $e_1,e_2,e_3$ are chosen to be spacelike.
Specifically, the frame is {\it pseudo-orthonormal}, i.e.,
$g(e_\alpha,e_\beta)=\text{diag}(+1,-1,-1,-1)=:o_{\alpha\beta}$.  Here
$g(u,v)\equiv u\cdot v$ denotes the scalar product of two vectors $u$
and $v$ defined by the Riemannian metric and $o_{\alpha\beta}$ are the
components of the Lorentz metric\footnote{... or, historically correct,
the Minkowski metric.}.

Three spacelike legs are not particularly practical if one wants to
investigate, say, electromagnetic or gravitational wave propagation.
Then, on the coordinate level, one introduces null coordinates $\sim
t\pm x$ (advanced and retarded ones).  Similarly, on the level
of the frame, one has two null legs $\sim e_0\pm e_1$. The two
remaining legs are untouched and stay spacelike. Such a {\it
half-null} frame leads immediately to the idea to transform also the
space-like legs to null legs. If one wants to uphold orthonormality,
this is only possible if one defines the new complex legs
$m,\overline{m}\sim e_2\pm i\,e_3$, with $i^2=-1$. This {\it
Newman-Penrose} null frame consists of two real and two complex null
vectors \cite{NP}. It turned out to be extremely useful for studying 
the properties of gravitational waves and of exact solutions of 
Einstein's field equation.

If one relaxes the constraint of orthonormality, one could also hope
that one is able to define a frame consisting of four {\it real} null
vectors.  In fact, such a frame was found by Coll \&
Morales\cite{Coll91}, and corresponding coordinates even earlier by
Derrick\cite{Derrick}, see also \cite{Coll85,Finkelstein,Saller,book}.
Amongst those frames, there are specifically the `symmetric null
frames' consisting of four legs none of which is geometrically
distinguished.  Such a null frame $f_\alpha$ can be constructed from
an orthonormal frame by a suitable non-degenerate linear
transformation.  Since this construction is not widely known, we will
present it in Sec.II in a hopefully easily accessible way. We will
study some of the properties of that frame and will visualize it by
means of a tetrahedron in 3-dimensional space. In Sec.III we will
introduce a real null {\it co\/}frame $\Phi^\alpha$ which turns out
{\it not} to be dual to the null frame $f_\alpha$.

Recently, Rovelli\cite{Rovelli}, in a framework related to the Global
Positioning System (GPS) \cite{Cornwall}, constructed four coordinates
$s^i$ in terms of which the components of the metric look null
symmetric, see also Coll\cite{Coll01}. In Sec.IV, we follow up these
ideas and show that Rovelli's coframe $ds^i$ is closely related to the
real coframe $\Phi^\alpha$.

\section{Frames consisting of four real null vectors}

At each point of 4-dimensional spacetime with coordinates $x^i$, here
$i,j\dots=0,1,2,3$, we have the 4-dimensional tangent vector space.
Four linearly independent vectors $e_\alpha$ constitute a basis or,
alternatively expressed, a {\it frame}. Dual to this frame is the {\it
coframe} $\vartheta^\beta$ which consists of 4 covectors (one-forms). 
One can decompose frame and coframe with respect to the local coordinate 
frame according to
\begin{equation}\label{decompose}
  e_\alpha=e^i{}_\alpha\,\partial_i\,,\qquad\vartheta^\beta
  =e_j{}^\beta\,dx^j\,,
\end{equation}
with the duality relations
\begin{equation}\label{duality}
  e_i{}^\alpha \,e^j{}_\alpha=\delta_i^j \qquad
  e_i{}^\alpha\,e^i{}_\beta=\delta^\alpha_\beta\,.
\end{equation}
The $e^i{}_\alpha$ are called frame (or tetrad) components.

Let the tangent vector space carry a metric $g$. Thereby we can define
a scalar product $g(u,v)\equiv u\cdot v$, where $u$ and $v$ are two
vectors.  Accordingly, the components of the metric with respect to
the frame $e_\alpha$ are determined by
$g_{\alpha\beta}:=g(e_\alpha,e_\beta)$.  Conversely, the metric can be
reconstructed from its components via
\begin{equation}\label{metric1}
  g=g_{\alpha\beta}\,\vartheta^\alpha\otimes\vartheta^\beta\,,
\end{equation}see Frankel\cite{Frankel}, for instance.

Traditionally, in relativity theory the vectors of an {\it 
orthonormal} frame are labeled by $0,1,2,3$, thus underlining the 
fundamental difference between $e_0$, which has positive length 
$g_{00}= g (e_0,e_0)=1$, and the $e_a$, $a=1,2,3$, which have negative 
lengths $g_{aa}= g (e_a,e_a)=-1$ (no summation).  In general, a vector 
$u$ is called {\it time-like} if $ g (u,u)>0$, {\it space-like} if 
$g(u,u)<0$, and {\it null} (or light-like) if $ g (u,u)=0$.  The 
components of the metric tensor with respect to the orthonormal frame 
$e_\alpha$ are then
\begin{equation}\label{eta}
  g_{\alpha\beta}\ \stackrel{*}{=}\ o_{\alpha\beta}\ :=
  \left(\begin{array}{crrr} 1 & 0 & 0 & 0\\ 0 &-1 & 0 & 0\\ 0 & 0 &-1
      & 0\\ 0 & 0 & 0 &-1
    \end{array}\right)=o^{\alpha\beta}\,.
\end{equation}
The star equal sign indicates that the corresponding equation is only
valid for specific frames, namely for orthonormal ones.

\subsection{Null frames}

Starting from an orthonormal frame $e_\alpha$ with respect to which
the metric has the standard form (\ref{eta}), we can build a new frame
$e_{\alpha'}=(l,n,e_{2'},e_{3'})$ by the linear transformation
\begin{equation}
l= {\frac{1}{\sqrt{2}}}(e_0 + e_1),\qquad n={\frac{1}{\sqrt{2}}}(e_0 - e_1),
\end{equation}
and $e_{2'}=e_{2}, e_{3'}=e_{3}$. In the kernel index method that we
are using, see Schouten\cite{Schouten}, the frames $e_{\alpha'}$ and
$e_\alpha$ are distinguished by the different type of indices. The
first two vectors of the new frame are null: $ g (l,l)= g (n,n)=0$.
Correspondingly, the metric with respect to this {\it half-null} frame
$e_{\alpha'}$ reads
\begin{equation}
  g_{\alpha'\beta'}\ \stackrel{*}{=} h_{\alpha'\beta'}
  :=\left(\begin{array}{ccrr} 0 & 1 & 0 & 0\\ 1 & 0 & 0 & 0\\ 0 & 0
      &-1 & 0\\ 0 & 0 & 0 &-1
    \end{array}\right)=h^{\alpha'\beta'}\,.\label{halfnull}
\end{equation}

Following Newman \& Penrose\cite{NP}, we can further
construct two more {\it null} vectors as the {\it complex} linear
combinations of $e_2$ and $e_3$:
\begin{equation}
m={\frac{1}{\sqrt{2}}}(e_2 + i\,e_3),\qquad\overline{m}={\frac{1}{\sqrt{2}}}
(e_2 - i\,e_3).
\end{equation}
Here $i$ is the imaginary unit and overbar means complex conjugation.
This transformation leads to the Lorentz metric in a {\it
Newman-Penrose} null frame $e_{\alpha''}=(l,n,m,\overline{m})$:
\begin{equation}
  g_{\alpha''\beta''}\ \stackrel{*}{=} n_{\alpha''\beta''} :=\left(
    \begin{array}{ccrr} 0 & 1 & 0 & 0\\ 1 & 0 & 0 & 0\\ 0 & 0 & 0
      &-1\\ 0 & 0 &-1 & 0
    \end{array}\right)=n^{\alpha''\beta''}\,.\label{NPnull}
\end{equation}
Such a frame is convenient for investigating the properties of
gravitational and electromagnetic waves.

In the Newman-Penrose frame, we have two real null legs, namely $l$
and $n$, and two complex ones, $m$ and $\overline{m}$. It may be
surprising to learn that it is also possible to define the {\it
  special null} frame which consists of four {\em real} null vectors.
We start from an orthonormal frame $e_\alpha$, with
$g(e_\alpha,e_\beta)=o_{\alpha\beta}$, and define the new frame
$f_{\widetilde{\alpha}}$ according to
\begin{eqnarray}\label{f2e}
f_{\widetilde{0}}&=&(\sqrt{3}\,e_0 + e_1 + e_2 + e_3)/2\,,\nonumber\\
f_{\widetilde{1}}&=&(\sqrt{3}\,e_0 + e_1 - e_2 - e_3)/2\,,\nonumber\\
f_{\widetilde{2}}&=&(\sqrt{3}\,e_0 - e_1 + e_2 - e_3)/2\,,\\
f_{\widetilde{3}}&=&(\sqrt{3}\,e_0 - e_1 - e_2 + e_3)/2\,.\nonumber
\end{eqnarray}
This can also be written as 
\begin{equation}\label{Saller1}
f_{\widetilde{\alpha}}=e_\beta\,F_{\widetilde{\alpha}}{}^\beta\,,
\qquad\text{with}\qquad
F_{\widetilde{\alpha}}{}^\beta=
 \frac{1}{2}\,\left(
            \begin{array}{rrrr}
               \sqrt{3}&  1&  1&  1 \\
               \sqrt{3}&  1& -1& -1 \\
               \sqrt{3}& -1&  1& -1 \\
               \sqrt{3}& -1& -1&  1
            \end{array}
            \right) \,.
\end{equation}
Since $ g (f_{\widetilde{\alpha}},f_{\widetilde{\alpha}})=0$ for all
$\widetilde{\alpha}$, the null frame consists solely of real
non-orthogonal null-vectors. The metric with respect to this frame
reads
\begin{equation}
  g_{{\widetilde{\alpha}}\widetilde{{\beta}}} \stackrel{*}{=}
  f_{\widetilde{\alpha}\widetilde{\beta}} := \left(
    \begin{array}{cccc} 0 & 1 & 1 & 1\\ 1 & 0 & 1 & 1\\ 1 & 1 & 0 &
      1\\ 1 & 1 & 1 & 0 \end{array}\right)\ne
  f^{\widetilde{\alpha}\widetilde{\beta}}\,.
\label{finkmat}
\end{equation}
This inequality means that the coframe dual to the frame
$f_{\widetilde{\alpha}}$ is {\em not null}.

\subsection{Properties  of real null frames}

The metric (\ref{finkmat}) looks completely symmetric in all its
components: Seemingly the time coordinate is not preferred in any
sense.  Nevertheless, Eq.(\ref{finkmat}) represents a truly Lorentzian
metric. Its determinant is $-3$ and the eigenvalues are readily
computed to be
\begin{equation}
  3, \quad -1, \quad -1, \quad -1,
\end{equation}
which shows that the metric (\ref{finkmat}) has, indeed, the correct
signature. The frame $f_{\widetilde{\beta}}$, by the linear
transformation $e_\alpha= f_{\widetilde{\beta}}\,
F_\alpha{}^{\widetilde{\beta}}$, can be transformed back to the
orthonormal frame $e_\alpha$. The matrix
$F_\alpha{}^{\widetilde{\beta}}$ is inverse to the matrix
$F_{\widetilde{\alpha}}{}^\beta$ in (\ref{Saller1}), i.e.,
$F_\alpha{}^{\widetilde{\beta}}
F_{\widetilde{\beta}}{}^\gamma=\delta_\alpha^\gamma$.

Provided the original orthonormal frame $e_\alpha$ is a coordinate or
{\it natural} frame, i.e., $e_\alpha=\delta_\alpha
^i\,\partial/\partial x^i$, then, because of (\ref{Saller1}) and
$F_{\widetilde{\alpha}}{}^\beta=\text{const.}$, the frame
$f_{\widetilde{\alpha}}$ is also natural. Accordingly, we can
introduce coordinates $\xi^i=(\tau,\xi,\eta,\zeta)$ such that (we drop
now the tilde)
\begin{equation}\label{natural}
  f_\alpha=\delta^i_\alpha\,\partial/\partial
  \xi^i\qquad\text{or}\qquad f^\alpha=\delta^\alpha_i\,d\xi^i\,.
\end{equation} Under those circumstances, the metric reads
\begin{eqnarray}
  g &=& f_{\alpha\beta}\, f^\alpha \otimes f^\beta\stackrel{*}{=}
  f_{ij}\, d\xi^i d\xi^j\nonumber\\ &=& 2 \left(
    d\tau\,d\xi+d\tau\,d\eta+d\tau\,d\zeta+
    d\xi\,d\eta+d\xi\,d\zeta+d\eta\,d\zeta \right)\,.
\label{finkmet}
\end{eqnarray}

There is a beautiful geometrical interpretation of the four null legs
of the frame $f_\alpha$, see Finkelstein\cite{Finkelstein}. In a {\em
  Minkowski spacetime}, let us consider the three-dimensional
spacelike hypersurface which is spanned by $(e_1,e_2,e_3)$.  The four
points which are defined by the spatial parts of the frame vectors
(\ref{f2e}), with coordinates $A=\frac{1}{2}\,(1,1,1)$,
$B=\frac{1}{2}\, (1,-1,-1)$, $C=\frac{1}{2}\,(-1,1,-1)$, and
$D=\frac{1}{2}\,(-1,-1,1)$, form a regular tetrahedron in the
3-dimensional subspace, see Fig.1. The vertices $A$, $B$, $C$, and $D$
are equidistant from the origin $O=(0,0,0)$; the distance is
$\sqrt{3}/2$. Correspondingly, all sides of this tetrahedron have equal
length, namely $\sqrt{2}$. If we now send, at the moment $t=0$, a
light pulse from the origin $O$, it reaches all four vertices of the
tetrahedron at $t=\sqrt{3}/2$.  Thus four light rays provide the
operational definition for the light-like frame $f_\alpha$ in
(\ref{f2e}).  In a Riemannian spacetime, we can choose Riemannian
normal coordinates.  Then, provided the tetrahedron is sufficiently
small, we will have an analogous interpretation.
\begin{figure}[htb]
  \centerline{{\psfig{figure=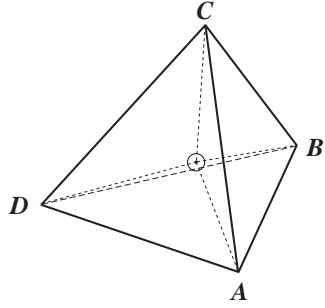, height=4cm}}}
  \bigskip\caption{A tetrahedron which represents the real symmetric
    null frame in 3-dimensional space. At time $t=0$, light is emitted
    in $O$. It reaches the points $A,B,C$, and $D$ at $t=\sqrt{3}/2$.
    The event $(0,O)$, together with the events $(\sqrt{3}/2,A)$,
    $(\sqrt{3}/2,B)$, etc., determine the four null vectors.}
\end{figure}

In the {\em special} real null frame (\ref{f2e}), all legs are
equivalent.  However, it is possible to apply a linear transformation
that keeps the zeros in the diagonal of (\ref{finkmat}) but changes
the off diagonal matrix elements such that the matrix remains
symmetric and non-degenerate. In this way, we find a {\em general}
real null frame.  Under such circumstances, the four real null vectors
$f_\alpha$ are no longer `indistinguishable'.

\section{Coframes consisting of four real null covectors}

The inverse of the matrix (\ref{finkmat}) is {\it not} the same
matrix, in contrast to the analogous cases (\ref{eta}),
(\ref{halfnull}), and (\ref{NPnull}). We rather find
\begin{equation}\label{finkinverse}
  f^{\alpha\beta} = \frac{1}{3}
\left(  \begin{array}{cccc} 
   -{2} & 1 & 1 & 1   \\ 
      1 & -{2} & 1 & 1\\ 
      1 & 1 &  -{2}& 1\\ 
      1 & 1 & 1 &  -{2}
\end{array}\right)\,.
\end{equation}
In other words, $f_{\alpha\alpha}=0$, but  $f^{\alpha\alpha}\ne 0$ 
(no summation over $\alpha$). However, we are able to find a new real 
null {\it co\/}frame by defining, in analogy 
to (\ref{f2e}),
\begin{eqnarray}\label{cofr}
  \Phi^{0'}&=&(\sqrt{3}\,\vartheta^0 + \vartheta^1 + \vartheta^2
     + \vartheta^3)/2,\nonumber\\ 
  \Phi^{1'}&=&(\sqrt{3}\,\vartheta^0 +\vartheta^1 - \vartheta^2 
     - \vartheta^3)/2,\nonumber\\ 
  \Phi^{2'}&=&(\sqrt{3}\,\vartheta^0 - \vartheta^1 + \vartheta^2 
     -\vartheta^3)/2,\\ \Phi^{3'}&=&(\sqrt{3}\,\vartheta^0 - \vartheta^1 
     -\vartheta^2 + \vartheta^3)/2.\nonumber
\end{eqnarray}
Accordingly, we have
\begin{equation}\label{nullcoframe1}
  \Phi^{\alpha'}={B}_\beta{}^{\alpha'}\,\vartheta^\beta\,,
                  \qquad\text{with}\qquad {B}_\beta{}^{\alpha'}
                = \frac{1}{2} \left(
  \begin{array}{cccc}
    \sqrt3 & \sqrt3 & \sqrt3 & \sqrt3\\
    1      &      1 &     -1 &     -1\\
    1      &     -1 &      1 &     -1\\
    1      &     -1 &     -1 &      1
  \end{array} \right)\,,
\end{equation}
where $B$ is the transpose of $F$ in (\ref{Saller1}): $B=F^{\rm T}$.  
Then, we obtain
\begin{equation}\label{nullcoframe2} 
  g^{\alpha'\beta'}= B_\mu{}^{\alpha'}B_\nu{}^{\beta'} \;{ o}^{\mu\nu}=
  (B^{\rm T}\,{o}\,B)^{\alpha'\beta'}= \Phi^{\alpha'\beta'}:=\left(
    \begin{array}{rrrr} 0& 1& 1& 1 \\ 1& 0& 1& 1 \\ 1& 1& 0& 1 \\ 1&
      1& 1& 0 \end{array} \right) \, .
\end{equation} 
Thus, $\Phi^{\alpha'}$ is a {\em special} null coframe 
with the contravariant metric components as given by 
(\ref{nullcoframe2}). 
\begin{figure}[htb]
  \centerline{{\psfig{figure=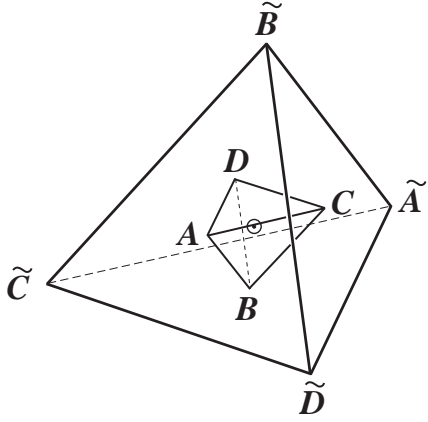,height=5.5cm}}}
  \bigskip\caption{The `outer' tetrahedron represents the real null
    coframe of (\ref{cofr}) with the contravariant metric components
    (\ref{nullcoframe2}). The `inner' tetrahedron is that of Fig.1.
    Note that the 3-vectors ${OA},{OB}$, etc. are perpendicular to the
    planes $\widetilde{B}\widetilde{C}\widetilde{D}$,
    $\widetilde{C}\widetilde{D}\widetilde{A}$, etc.}
\end{figure}

The visualization of the new null coframe leads also to a tetrahedron
in 3-space which is ``dual'' to the one of Fig.1. We depicted that in
Fig.2. The vertices of the coframe tetrahedron are 
$\widetilde{A}=\!-\frac{3}{2}\,(1,1,1)$,
$\widetilde{B}=\!-\frac{3}{2}\, (1,-1,-1)$,
$\widetilde{C}=\!-\frac{3}{2}\,(-1,1,-1)$, and
$\widetilde{D}=\!-\frac{3}{2}\,(-1,-1,1)$. Its sides have a length of 
$3\sqrt{2}$. The barycenters of the four triangles of the tetrahedron 
are the vertices $A,B,C,D$ of the tetrahedron of Fig.1.

According to the construction, the 3-vectors $OA,\,OB,\, OC$, and
$OD$, representing light rays, are perpendicular to the triangles
$\widetilde{B}\widetilde{C}\widetilde{D},\,
\widetilde{C}\widetilde{D}\widetilde{A},\,
\widetilde{D}\widetilde{A}\widetilde{B}$, and
$\widetilde{A}\widetilde{B}\widetilde{C}$, respectively.  Therefore,
at the points $A$ etc., the planes
$\widetilde{B}\widetilde{C}\widetilde{D}$ etc.\ represent the wave
fronts of the light.  Together with the parallel planes through
$O=(0,0,0)$, the planes $\widetilde{B}\widetilde{C}\widetilde{D}$
etc.\ symbolize covectors that are algebraically dual to the vectors
$OA$ etc. The interior product of the vectors with the corresponding
covectors is always $1$. In this sense, the outer tetrahedron is dual
to the inner one.

The most general null coframe has the form
\begin{equation}
\Phi^{\overline\alpha}
    ={B}_\beta{}^{\overline\alpha}\,\vartheta^\beta\,,
        \qquad \text{with} \qquad {B}_\beta{}^{\overline\alpha}
    =\left(
    \begin{array}{llll} 
      \alpha_0& \alpha_1& \alpha_2& \alpha_3 \\ 
      \beta_0& \beta_1& \beta_2& \beta_3     \\ 
      \gamma_0& \gamma_1& \gamma_2& \gamma_3 \\ 
      \delta_0& \delta_1& \delta_2& \delta_3 \\ 
    \end{array}\right)\, ,                                \label{m3}
\end{equation}
where $\alpha^2:=\alpha_\mu\alpha_\nu\,o^{\mu\nu}= \beta^2=\gamma^2=
\delta^2=0$ and $\det B\ne 0$.  The corresponding contravariant
components of the metric read
\begin{equation}
   g^{\overline\alpha\overline\beta}
    =\left(\begin{array}{cccc} 
      0 & \alpha\cdot\beta& \alpha\cdot\gamma& \alpha\cdot\delta \\ 
     \beta\cdot\alpha& 0 & \beta\cdot\gamma& \beta\cdot\delta \\ 
     \gamma\cdot\alpha& \gamma\cdot\beta& 0 & \gamma\cdot\delta \\ 
     \delta\cdot\alpha& \delta\cdot\beta& \delta\cdot\gamma& 0 \\ 
           \end{array}\right)   \, .                       \label{m4}
\end{equation}
For ${B}_\beta{}^{\overline\alpha}= {B}_\beta{}^{\alpha'}$, the general 
expression for $g^{\overline\alpha\overline\beta}$ reduces to the special 
form (\ref{nullcoframe2}).

\section{Rovelli's construction and real null coframes}

Let us eventually turn to Rovelli's paper \cite{Rovelli} which prompted 
our remarks in the first place. The 4-velocity of a massive particle 
moving in a (flat) Minkowski space $M_4$, expressed in the standard 
{\it inertial} coordinates $x^i$, has the form 
\begin{equation}                                           \label{e1}
  W^i:=\frac{dx^i}{ds}=\frac{1}{\sqrt{1-v^2}}\,\left(1,v^a\right)\, ,
\end{equation}
where $v^a:= dx^a/dt=v\,n^a$ represents the 3-velocity and $n^a$ a
unit 3-vector.  We assume $v$ to be constant. The 4-vector $W^i$ is
normalized, i.e., $W^2:= W\cdot W=1$. For constant $W^i$, the 
particle's trajectory is of the simple form 
\begin{equation}                                          \label{e2} 
x^i(s)=s\, W^i\, .
\end{equation}

Suppose there is an observer at some point $P$ with coordinates 
$x^i=X^i$ who is able to detect light signals emitted from the moving 
particle. The observer's light cone is described by the equation 
\begin{equation}                                          \label{e3} 
(x-X)^2=0\,.                                                
\end{equation}
The condition that the particle's trajectory intersects this light 
cone is obtained by solving the system of equations (\ref{e2}) and 
(\ref{e3}). Substituting (\ref{e2}) into (\ref{e3}), 
we obtain a quadratic equation for $s$, namely $s^2-2s\,W\cdot X+X^2=0$, 
which yields
\begin{equation}                                          \label{e4} 
s=W\cdot X-\sqrt{ (W\cdot X)^2-X^2 }\, .
\end{equation}
The minus sign given in front of the square root refers to the past 
light cone, whereas the plus sign is suppressed since it refers to the 
future light cone. We should note that the result in 
(\ref{e4}) is correct for {\it any\/} choice of the unit 
vector $n$ in $W$. 
\begin{figure}[htb]
\centerline{{\psfig{figure=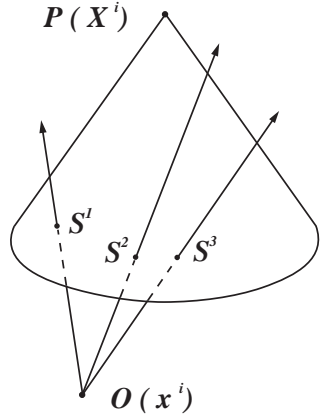,height=5.5cm}}}
\bigskip \caption{This is a $1+2$-dimensional spacetime diagram, i.e., 
one spatial coordinate is suppressed. The diagram represents Rovelli's
satellites which supply the `distance' coordinates $s^1, s^2, s^3$
for the event $P(X^i)$. In $1+3$ dimensions, we have, of course, 
{\it four\/} observers.}
\end{figure}

If we consider a set of four particles, see Fig.3, in which one space 
dimension is suppressed and thus only three particles are visible, 
then Eqs.(\ref{e1}) and (\ref{e4}) are replaced by 
equations of the same type, 
\begin{equation} 
  W^{i'}_k=\frac{1}{\sqrt{1-(v^{i'})^2}} 
  \left(1,v^{i'}n^{i'}_a\right)\, ,\quad s^{i'}=W^{i'}\cdot 
  X-\sqrt{(W^{i'}\cdot X)^2-X^2}\, . 
\end{equation} 
Here $W^{i'}\cdot X:=W^{i'}_k\,X^k$ and $i'=0,1,2,3$ is a label for 
different particles. Note that the four 3-velocities $v^{i'}$ are in 
general different from each other, in contrast to the original Rovelli 
construction where they are all equal.  This generalization is 
physically quite natural, and it will be important in our 
considerations. In the next step, following Rovelli, we introduce 
$s^{i'}$ as new coordinates on the $M_4$. The Jacobian matrix of the 
transformation $x^k\to s^{i'}$ at the observer's point $P(X)$ is given 
by 
\begin{equation} 
E_k{}^{i'} =\frac{\partial s^{i'}}{ \partial x^ k }(x=X) 
       =\frac{\partial s^{i'}}{\partial X^ k } 
       =W^ {i'} _ k -\frac{W^ {i'} _ k (W^ {i'} \cdot X)-X_ k}   
                 {\sqrt{(W^ {i'} \cdot X)^2-X^2} }\, .  
\end{equation}
The contravariant components of the metric in the new coordinates are
\begin{eqnarray}
  g^{ {i'} {j'} }&&=E_ k{}^ {i'} E_l{}^{j'} \, o^{ k l } \nonumber\\ 
  &&=W^ {i'}\cdot W^ {j'} -\left[\frac {(W^ {i'} \cdot W^ {j'} )
     (W^{j'}\cdot X)-(W^{i'}\cdot X)} {\sqrt{(W^{j'}\cdot X)^2-X^2}} 
     +( {i'} \leftrightarrow {j'} )\right]\nonumber\\ 
  &&\quad+\frac{(W^{i'}\cdot W^{j'} )(W^{i'}\cdot X)(W^ {j'}\cdot X)
     -(W^{i'}\cdot X)^2-(W^{j'}\cdot X)^2+X^2}{ \sqrt{(W^{i'}
      \cdot X)^2-X^2} \sqrt{(W^{j'} \cdot X)^2-X^2 } }\, . 
\end{eqnarray}
A direct calculation for $ {j'} = {i'} $, by using $(W^ {i'} )^2=1$,
leads to \begin{equation}                                \label{diag} 
  g^{{i'}{i'}}=\frac{ \partial s^{i'}}{\partial x^k}
  \frac{ \partial s^{i'}}{\partial x^l}\,o^{kl} =0 
  \qquad  ({i'},j' =0,1,2,3)\, .                      
\end{equation} 
Note that this result is valid for {\it any\/} choice of the unit
vectors $n^ {i'}$ in $W^ {i'} $. The components $g^{{i'}{j'}}$,
for ${i'}\ne {j'}$, depend only on the scalar products $W^{i'}\cdot
W^{j'}$ and $W^{i'}\cdot X$. The surfaces $s^{i'}=const$ are null
hypersurfaces.  We now drop the primes for simplicity.

In {\it frame} (\ref{f2e}), we have $g_{\alpha\alpha }=0$, but 
$g^{  \alpha\alpha }\ne 0$, in contrast to equation (\ref{diag}).
However, for the general real null coframe of (\ref{m3}), we have the 
desired $g^{\alpha\alpha}=0$. Therefore it is evident that the Rovelli 
coordinates are closely related to this general real null coframe. 

Rovelli's coframe $E^\alpha=E_k{}^\alpha\,\vartheta^k$ belongs to the 
class of general null coframes (\ref{m3}), hence it can be reduced to 
our special null coframe (\ref{nullcoframe1}) by demanding 
\begin{equation}
  E{_k}{}^\alpha= B{_k}{}^\alpha                         \label{m5}
\end{equation}
or, more explicitly,
\begin{eqnarray}
&& W^0_k-\frac{W^0_k(W^0\cdot X)-X_k}{\sqrt{(W^0\cdot X)^2-X^2}}
   =  (\sqrt{3},\>1,\>1,\>1)/2  \, ,             \nonumber\\
&&  W^1_k-\frac{W^1_k(W^1\cdot X)-X_k}{\sqrt{(W^1\cdot X)^2-X^2}}
   = (\sqrt{3},1,-1,-1)/2\, ,               \nonumber\\
&& W^2_k-\frac{W^2_k(W^2\cdot X)-X_k}{\sqrt{(W^2\cdot X)^2-X^2}}
   =  (\sqrt{3},-1,1,-1)/2 \, ,               \nonumber\\
&& W^3_k-\frac{W^3_k(W^3\cdot X)-X_k}{\sqrt{(W^3\cdot X)^2-X^2}}
   =  (\sqrt{3},-1,-1,1)/2 \, .                          \label{m6}
\end{eqnarray}

We shall now find a particular solution of these equations for
$W^\alpha_k$, for a given position $X$ of the observer. Without loss
of generality, we can choose the inertial coordinates $x^i$ in such a
way that
\begin{equation}
X^i=({\cal T},{\cal X},0,0)\, ,
           \qquad {\cal T}> 0\,,\,\,{\cal X}\ne 0\, .   \label{m7}
\end{equation} 
In that case, by multiplying Eqs.(\ref{m6}) with $X^k$, we find
\begin{eqnarray}
  &&(W^0\cdot X)-\sqrt{(W^0\cdot X)^2-X^2} =(\sqrt{3}{\cal T}
    +{\cal X})/2 =: b^0 \, ,\nonumber\\ 
  &&(W^1\cdot X)-\sqrt{(W^1\cdot X)^2-X^2} =(\sqrt{3}{\cal T}
    +{\cal X})/2 =: b^1 \, ,\nonumber\\ 
  &&(W^2\cdot X)-\sqrt{(W^2\cdot X)^2-X^2} =(\sqrt{3}{\cal T}
    -{\cal X})/2 =: b^2 \, ,\nonumber\\ 
  &&(W^3\cdot X)-\sqrt{(W^3\cdot X)^2-X^2} =(\sqrt{3}{\cal T}
    -{\cal X})/2 =: b^3 \, .                              \label{m8}
\end{eqnarray}
Since $b^0=b^1$ and $b^2=b^3$, it follows that
\begin{equation}
W^0\cdot X=W^1\cdot X\, ,\qquad  W^2\cdot X=W^3\cdot X\, .
\end{equation}
This result combined with Eqs.(\ref{m6}) leads to
\begin{eqnarray}
&&W^0_0=W^1_0 \, ,\quad W^0_1=W^1_1  \, , \quad 
  W^0_2=-W^1_2\, ,\quad W^0_3=-W^1_3 \, ,      \nonumber\\ 
&&W^2_0=W^3_0 \, ,\quad W^2_1=W^3_1  \, , \quad
  W^2_2=-W^3_2\, ,\quad W^2_3=-W^3_3 \, .      
\end{eqnarray}
A simple form of $W^\alpha_k$, consistent with these conditions, is
given by
\begin{eqnarray}
&&W^0_k=\frac{1}{\sqrt{1-v^2}}\left(1,-v\,n^{0a}\right)\, ,\nonumber\\
&&W^1_k=\frac{1}{\sqrt{1-v^2}}\left(1,-v\,n^{1a}\right)\, ,\nonumber\\
&&W^2_k=\frac{1}{\sqrt{1-u^2}}\left(1,-u\,n^{2a}\right)\, ,\nonumber\\
&&W^3_k=\frac{1}{\sqrt{1-u^2}}\left(1,-u\,n^{3a}\right)\, ,  \label{m9}
\end{eqnarray}
where $v$ and $u$ are undetermined velocities and the $n^{\alpha}$
are the 3-dimensional unit vectors of (\ref{f2e}):
\begin{equation}
\begin{array}{ll}
n^{0a}=(1,1,1)/\sqrt{3}\, ,\quad
                  & n^{1a}=(1,-1,-1)/\sqrt{3}\, ,\\
n^{2a}=(-1,1,-1)/\sqrt{3}\, ,\quad
                 & n^{3a}=(-1,-1,1)/\sqrt{3}\, .
\end{array}
\end{equation}

We now use Eqs.(\ref{m8}) in order to express $W^\alpha\cdot X$ in
terms of $X^i$:
\begin{equation}\label{last}
W^\alpha\cdot X=\frac{(b^\alpha)^2+X^2}{2b^\alpha}=:h^\alpha \, .
\end{equation}
Then we substitute $W^\alpha$ from Eqs.(\ref{m9}) into (\ref{last}), 
and obtain (only) two independent equations:
\begin{eqnarray}
  &&\frac{1}{\sqrt{1-v^2}} \left({\cal T}-\frac{1}{\sqrt{3}}v\,{\cal
      X}\right) =h^0\, , \nonumber\\ 
  &&\frac{1}{\sqrt{1-u^2}}\left({\cal T}+\frac{1}{\sqrt{3}}u\,{\cal
      X}\right) =h^2\, .\                                \label{m10}
\end{eqnarray}
Taking into account the explicit expressions for $h^0$ and $h^2$, these 
equations can be written in the simple form
\begin{eqnarray}
&&a(v)\,{\cal T}^2+b(v)\,{\cal T}{\cal X}+c(v)\,{\cal X}^2=0\,,\nonumber\\
&&a(u)\,{\cal T}^2-b(u)\,{\cal T}{\cal X}+c(u)\,{\cal X}^2=0\,,\label{j1}
\end{eqnarray}
with
\begin{eqnarray}
&&a(w):= 4\sqrt{3}-7\sqrt{1-w^2}\, ,\nonumber\\
&&b(w):=4(1-w)-2\sqrt{3}\sqrt{1-w^2}\, ,\nonumber\\
&&c(w):=3\sqrt{1-w^2}-4w/\sqrt{3}\, ,
\end{eqnarray}
for $w=v$ or $u$. We demand that Eqs.(\ref{j1}) have real solutions
for ${\cal T}$ and ${\cal X}$. It implies the following conditions on
the velocities:
\begin{equation} 
\Delta(w):=b^2(w)-4a(w)\,c(w)\ge 0\, .
\end{equation}
The calculations presented in Fig.4 show the allowed values for the
velocities $w$.

\begin{figure}[htb]
  \centerline{{\psfig{figure=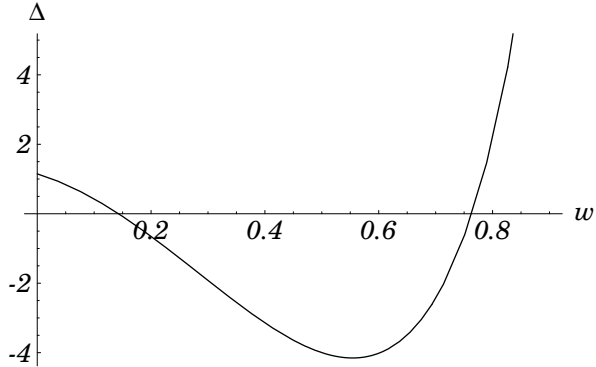, height=5cm}}}
  \bigskip\caption{The allowed values of $w$ are determined by
    $\Delta(w)\ge 0$.}
\end{figure}

\noindent A particular example of an acceptable set of parameters is given by 
$\{{\cal T}=1/\sqrt{3},\, {\cal X}=4,\,v=0.77,\, u=0.83\}$.  In
general, for all sets of parameters $\{{\cal T},{\cal X},v, u\}$ which
satisfy the requirements above, Rovelli's null coframe reduces to our
special null coframe (\ref{nullcoframe1}).

In conclusion, in this paper we introduced and discussed the special
real null coframe (\ref{nullcoframe1}) whereupon the meaning of
Rovelli's coordinates in Minkowski spacetime becomes clearer. In the
Riemannian space of general relativity, Rovelli's coordinates are
related to the class of general null coframes (\ref{m3}).

\subsection*{Acknowledgments} We are grateful to David Finkelstein 
for discussions on the null frame and to Bartolom\'e Coll (Paris) for
informing us about Derrick's and his work.  Furthermore, we thank the
DAAD, Bonn (J.G.\ and M.B.) and the Humboldt Foundation, Bonn (Y.N.O.)
for financial support. J.G.\ and M.B.\ would like to thank F.W.\ Hehl
for the warm hospitality extended to them during their stay at the
University of Cologne.

\end{document}